# Palm-sized, vibration-insensitive and vacuum-free all-fiber-photonic module for $10^{-14}$-level stabilization of CW lasers and frequency combs

Igju Jeon[1], Changmin Ahn[1], Chankyu Kim[1], Seongmin Park[1], Wonju Jeon[1], Lingze Duan[2] and Jungwon Kim[1,*]

[1]Department of Mechanical Engineering, Korea Advanced Institute of Science and Technology (KAIST), Daejeon 34141, Korea
[2]Department of Physics and Astronomy, University of Alabama in Huntsville, AL 35899, USA
(*Electronic mail: jungwon.kim@kaist.ac.kr)

**Compact and robust frequency-stabilized laser sources are critical for a variety of fields that require stable frequency standards, including field spectroscopy, radio astronomy, microwave generation, and geophysical monitoring. In this work, we applied a simple and compact fiber ring-resonator configuration that can stabilize both a continuous-wave laser and a self-referenced optical frequency comb to a vibration-insensitive optical fiber delay-line. We could achieve a thermal-noise-limited frequency noise level in the 10 Hz – 1 kHz offset frequency range for both the continuous-wave laser and the optical frequency comb with the minimal frequency instability of $2.7\times10^{-14}$ at 0.03-s and $2.6\times10^{-14}$ at 0.01-s averaging time, respectively, in non-vacuum condition. The optical fiber spool, working as a delay reference, is designed to be insensitive to external vibration, with a vibration sensitivity of sub-$10^{-10}$ [1/g] and volume of 32 mL. Finally, the ring-resonator setup is packaged in a palm-sized aluminum case with 171-mL volume with a vibration-insensitive spool, as well as an even smaller 97-mL-volume case with an ultra-compact 9-mL miniaturized fiber spool.**

---

## I.    Introduction

Ultra-stable single-wavelength lasers and optical frequency combs are critical to many fields that require extremely stable frequency references, such as optical lattice clocks [1,2], gravitational wave detection [3], precision spectroscopy [4,5] and redefinition of SI units [6]. They can be also used to generate ultra-stable microwaves through optical frequency division (OFD) [7,8], which can improve the performance of various microwave photonic applications including radio astronomy and radar systems [9,10]. In recent years, from earthquake detection [11] to field-deployed spectroscopy [12], there is a growing demand for more compact, robust and portable ultra-stable laser systems operating in non-laboratory environment.

There have been several approaches for realizing compact-size laser stabilization platforms, where most representative examples include compact ultra-low expansion (ULE) cavities [13–17], integrated chip-scale resonators [18,19], bulk dielectric whispering-gallery-mode (WGM) resonators [20,21], stimulated Brillouin scattering (SBS) [22,23] and optical fiber delay-lines [24-31]. The lasers stabilized by compact vacuum-gap Fabry-Pérot (FP) cavities are recently reported to reach $10^{-15}$–level frequency instability at an integration time of 0.1 s [13,14] and 0.01s [16]. While the platform provides ultra-low frequency noise stabilization, the system is generally highly complex and requires isolation from the environment using a high-vacuum system and multiple layers of thermal shields. In case of compact WGM resonator-based laser stabilization, the state-of-the-art long-term frequency instability is on the order of $10^{-14}$ at 0.1 s when using $MgF_2$ with ultra-high Q factor of 2 billions [21]. Although the resonator itself offers a compact structure, laser coupling is complex due to its prismatic coupling system and is susceptible to environment perturbations such as vibration, thermal drift and acoustic noise, and again requiring a bulky vibration isolation platform, thermal shields and vacuum. While the integrated chip-scale resonators [18,19] do not require vacuum system, the minimum frequency instability of stabilized laser is limited to $10^{-13}$-level at near 0.01 s. Recently, optical fiber-based SBS process could generate narrow linewidth optical signals [22,23] with frequency instability of $10^{-13}$ at less than 0.1 s.

Alternatively, an optical fiber delay-line can be used to stabilize the laser using a self-heterodyne method [24-29]. Because the platform is a fully fiber optic system, the system is alignment-free, robust, and consists of off-the-shelf fiber-coupled components. Moreover, the platform can be directly applied to the stabilization of not only CW lasers [24] but also mode-locked fiber laser combs [25-27] and Kerr micro-combs (from tens-GHz [28] to hundreds-GHz repetition rates [29]). Recently, the repetition rate of Kerr frequency comb was stabilized to the $10^{-13}$-level frequency instability at 0.4 s, where the packaging size was mostly limited by the size of the acousto-optic frequency shifter (AOFS) [28]. Although the development of gyroscope fiber optic coil winding technology makes it possible to make a 1-km-long fiber spool compact in size (~7-cm diameter), the necessity of modulators such as AOFS limited the overall size of the system. In this paper, we show how to stabilize a 1550-nm CW fiber laser as well as comb-lines of a self-referenced optical frequency comb to $10^{-14}$-level frequency instability without the use of a vacuum system using an all-fiber optical delay-line-based, modulator-free ring-resonator configuration, which is well suited for field applications outside of a laboratory environment. This configuration can offer large Q factor ($4.5\times10^9$) using a 100-m-long polarization-maintaining (PM) fiber. While a fiber ring resonator based on the PDH method was recently demonstrated to stabilize a 1550-nm CW laser frequency [30], in this work, the generation of an error signal was accomplished using a simpler balanced photodetection [31] without the use of a modulator. This also allowed for the stabilized frequency noise level to reach the thermal noise limit of the 100-m-long fiber. Furthermore, in order to suppress the length fluctuation of the fiber delay due to external vibrations, the shape of the fiber spool mount is specially designed to have vibration sensitivity of ~10⁻¹⁰ [1/g] level. Finally, the optical setup was packaged in a palm-sized case and tested.

## II. Fiber ring-resonator

The PM fiber-ring resonator is composed of four components: an unbalanced optical fiber coupler, an optical isolator, an optical fiber delay-line, and a balanced 50:50 optical fiber coupler for balancing (see Fig. 1(a)). Here we employed a 99:1 coupler to store the majority of input optical power in the resonator, the optical isolator to prevent Brillouin backscattering [30], and the optical fiber delay-line to increase the finesse and Q factor. The resonator is constructed in PM configuration in order to maximize the signal-to-noise ratio of the output interference optical signal by maintaining the polarization state in the fiber ring resonator. To avoid the intensity-noise-coupled error signal, we additionally used a 50/50 coupler to divide the laser output by two, one of which goes through the resonator and the other is used for power level subtraction. The balanced photodetection of the two ports can effectively eliminate the conversion of laser intensity noise into frequency noise [31]. The transmittance of the ring resonator without balancing is expressed as

$$\left|\frac{E_{o,r}}{E_i}\right|^2 = \left|\frac{\left[t - \sqrt{\alpha} \cdot \exp\left(\frac{2\pi f n l}{c}\right)\right]}{\left[1 - \sqrt{\alpha} \cdot t \cdot \exp\left(\frac{2\pi f n l}{c}\right)\right]}\right|^2 \quad (1)$$

where $E_i$ and $E_{o,r}$ are the electric field of input laser and output port of the ring resonator, respectively, $t$ is the square root of the coupling ratio, $l$ is the length of delay, $\alpha$ is the optical attenuation along the resonator, $f$ is the optical frequency of the incident laser, $n$ is the refractive index of the fiber, and $c$ is the speed of light. The transmittance function forms periodic dips spaced by the free spectral range (FSR) where the optical frequency can be discriminated with the sensitivity proportional to the Q factor of the resonator [curve (i) in Fig. 1(b)]. For example, for the system with $t^2$=0.99, $l$=100 m, and internal loss of 0.5 dB, the resonator has the FSR of 2 MHz and Q factor of 4.5×10⁹. The use of a ring resonator can significantly reduce the necessary fiber length: for example, to fulfill the same Q factor (4.5×10⁹), the simple delay-line requires a 2.2 km-long fiber delay, which is 22 times longer than the ring resonator case. The locking point is determined by the level of the balancing port with constant transmittance as $\left|E_{o,b} / E_i\right|^2$ which is shown as curve (ii) in Fig. 1(b), where $E_{o,b}$ is the electric field of the output balancing port.

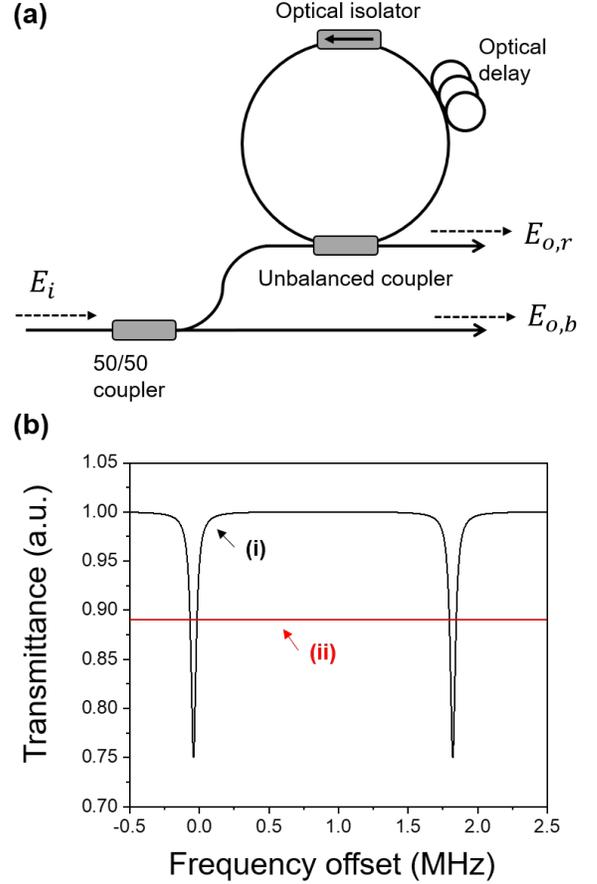

Fig. 1. (a) Configuration of the fiber ring resonator with balancing port. (b) Theoretical transmittance of (i) ring resonator port and (ii) balancing port. The transmittance function for the ring resonator port is calculated when the fiber delay is 100 m long and the cavity loss is 0.5 dB.

## III. Vibration-insensitive fiber spool design

Various mechanical perturbations such as acoustic noise, vibration, and thermal drift coupled to the optical references cause optical length fluctuation followed by the frequency fluctuation of the stabilized laser. To reduce the vibration-induced frequency noise, passive vibration isolation platform or active vibration control [32] can be used. However, such systems are generally bulky and heavy, which compromises the compactness and light weight of optical fiber-based stabilization method. Therefore, there was previous research on the design and implementation of vibration-insensitive fiber spools [33], which recently showed down to 8×10⁻¹¹ [1/g] vibration sensitivity for a 1-km-long single-mode fiber delay-line [34], where the vibration sensitivity of the mechanical length reference is defined as the length

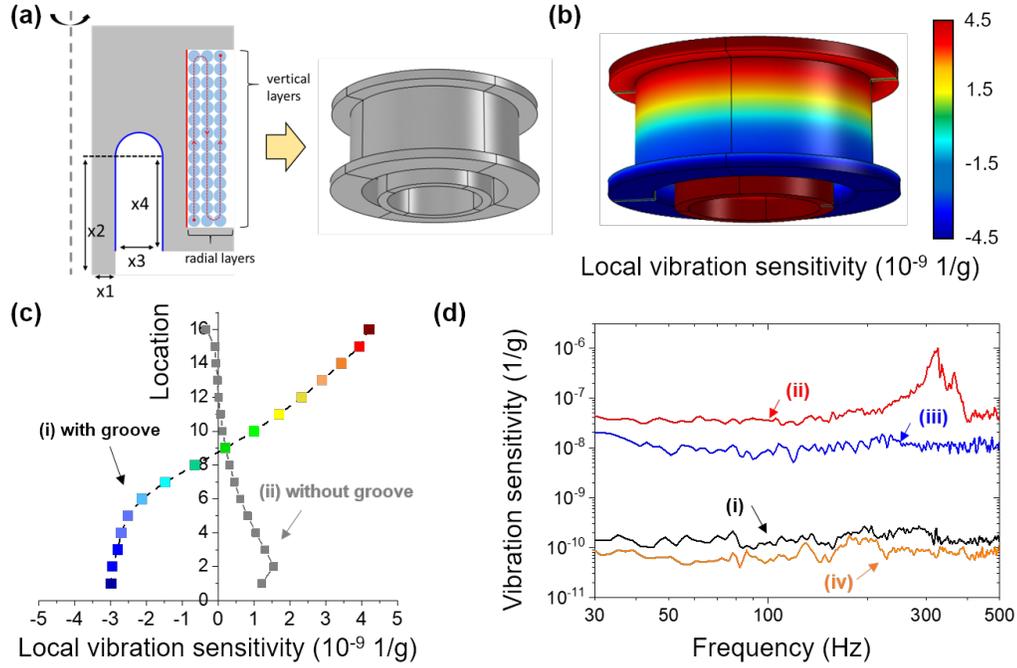

Fig. 2. Vibration-insensitive spool design and measured vibration sensitivity of various spools. (a) Design of a vibration-insensitive optical fiber spool. The fiber (cross section as small circles) is wound in the direction marked by the red dashed-line. (b) Visualization of local vibration sensitivity with vibration at the bottom of the spool. (c) The local vibration sensitivity (i) with groove and (ii) without groove. (d) The vibration sensitivity of various types of 100-m PMF fiber spools and a 1-km SMF fiber spool. (i) designed vibration-insensitive spool; (ii) typical 15-cm diameter plastic spool; (iii) miniaturized (26-mm diameter) aluminum spool without grooves; (iv) vibration-insensitive 1-km SMF spool.

expansion ratio ($dl/l$) induced by the vibration with unit acceleration of g.

In this work, we designed a compact, lightweight and simple vibration-insensitive spool for the 100-m-long PM fiber with $\sim10^{-10}$ [1/g] vibration sensitivity (with the minimum value of $9.4 \times 10^{-11}$ [1/g]), which is made with aluminum and has a volume of only 32 mL and weight of 41 g. The design of the spool is sketched in Fig. 2(a). The fiber is wound from bottom to top and top to bottom alternatively one by one (radial) layer, then stacked to the radial direction. To remove the relative movement between radial layers, which can cause collapse of the spool, a small amount of epoxy was used to fix the fiber when winding each layer. When the vibration is applied to the cylinder where fiber is wound, the fiber expands or contracts according to its winding position. The groove (marked in blue) in the design makes some winding parts (marked in red) of the spool expand and the other parts contract, which makes the net length fluctuation of fiber near zero. While the fiber wound on a cylinder with no groove expands or contracts in all positions, if a groove is present, fiber wound on upper and lower part can expand and contract in the opposite direction. Note that the groove is designed in a round shape in order to avoid the stress concentration that can occur when groove is in an angled shape. Using the COMSOL Multiphysics FEM simulation tool, four parameters (x1, x2, x3, and x4) shown in Fig. 2(a) are tuned to balance out the expansion and contraction and make the overall vibration sensitivity zero. The objective function for optimization considers both the vibration sensitivity itself and the potential sensitivity error caused by machining errors, and the function value needs to be minimized for optimization. The objective function is

$$O = \left(\max\left(\log\left|f\right| + 12, 0\right)\right)^2 + \log\left\|\frac{\partial f}{\partial x}\right\|_2 \tag{2}$$

where $f$ is the vibration sensitivity of the designed spool and $x$ is the vector of (x1, x2, x3, x4). We designed the first term of the objective function to reduce the vibration sensitivity as much as possible down to the $10^{-12}$ [1/g] level. If the value of vibration sensitivity is smaller than $10^{-12}$ [1/g], $\left(\max\left(\log\left|f\right| + 12, 0\right)\right)^2$ term will produce zero. This will induce the optimization procedure to focus more on minimizing the second term (i.e., the first derivative of the vibration sensitivity with respect to the four parameters), which is used to minimize the error caused by the machining error of the spool. As a result of the optimization, the spool is designed to have a vibration sensitivity as $5 \times 10^{-11}$ [1/g]. The first derivative of the vibration sensitivity $\left\|\frac{\partial f}{\partial x}\right\|_2$ is optimized to $2.5 \times 10^{-7}$ [1/(g·m)], where the multiplication with machining error (0.03 mm) results in $7.5 \times 10^{-12}$ [1/g], which is almost ten times lower than the objective vibration sensitivity. The local vibration sensitivity of the designed spool is presented graphically in Fig. 2(b). Fig. 2(c) shows the results of FEM simulations of the local vibration sensitivity, which is the vibration sensitivity of fiber coiled on a certain height of the spool. When a cylinder has no groove, fiber wound on all positions commonly expands or contracts; if a groove is present, fiber wound on the upper and lower part expands and contracts in the opposite direction.

To validate the optimization, we assessed the vibration sensitivity of the designed vibration-insensitive 100-m PM fiber spool. We applied the vibration to the in-loop ring resonator shown in Fig. 3(a), then measured the frequency noise of the stabilized laser using the out-of-loop frequency noise measurement setup depicted in Fig. 3(c). The measured frequency noise is converted to the length fluctuation spectrum, which is then divided by the vibration spectrum to obtain the vibration sensitivity spectrum. For comparison, we also measured the vibration sensitivity of two other 100-m PM fiber spools, which are wound on a 15-cm-diameter plastic spool and a miniaturized 26-mm-diameter aluminum spool. The measured vibration sensitivities of these spools are shown in Fig. 2(d). The vibration sensitivity of the optimized spool is shown as curve (i), which shows a flat spectrum along the broadband input vibration frequency from 30 Hz to 500 Hz, with $\sim 10^{-10}$ [1/g] level (with minimum value of $9.4 \times 10^{-11}$ [1/g]). It is slightly higher than the simulation result, and the imperfect winding procedure and unequal transmission from the bottom of the spool to the fiber appear to be the causes of the error. The vibration sensitivity of the typical plastic spool is shown as curve (ii) with vibration sensitivity of $4 \times 10^{-8}$ [1/g], which is $\sim 400$ times higher than the designed vibration-insensitive spool. The miniaturized aluminum spool with a diameter of 26 mm (48 mm when including the wound fiber) and a height of 5 mm has the vibration sensitivity of $10^{-8}$ [1/g] [curve (iii) in Fig. 2(d)]. Additionally, vibration-insensitive 1-km single-mode fiber spool (that is used for the out-of-loop noise measurement, see Fig. 3(b)) is also designed in the same way to have vibration sensitivity of $2.0 \times 10^{-12}$ [1/g] calculated from simulation in order to prevent the impact of vibrations when measuring the frequency noise. The vibration sensitivity of the spool is shown as curve (iv) in Fig. 2(d), with the level of under $10^{-10}$ [1/g] (with minimum value of $3.9 \times 10^{-11}$ [1/g]). Note that the slightly lower vibration sensitivity of the 1-km SMF spool compared to the 100-m PMF spool also comes from the imperfect winding procedure. Ideally, the vibration sensitivity may not be influenced by the fiber length, since it is primarily determined by the mechanical strain of the bulk fiber. While we modeled the optical fiber as a bulk fiber in the simulation, in reality, the fiber is wound layer by layer with epoxy to fix it. The amount of epoxy is not perfectly uniform, causing the difference in material property and dimension between simulation and experiment.

## IV. Experiments

Fig. 3(a) shows the schematic of the CW laser stabilization system based on the fiber ring resonator. A 1550-nm PM erbium-doped fiber laser (ETH-20-1550.12-PZ10B, Orbits Lightwave, Inc) is used as the laser source. The laser output goes through an acousto-optic frequency shifter 1 ($AOFS_1$) which works as an extra-cavity fast frequency modulator. Then 50/50 coupler divides the laser output into two, where one is applied to the ring resonator and the other is directly applied to one of the input ports of the balanced photodetector. Frequency noise for each environment is compared after the fiber reference, which includes a ring resonator and balancing ports, when put in a low-vacuum chamber ($\sim$1-torr pressure), an airtight chamber, or an unsealed open chamber. The output of the balanced photodetector, which works as an error signal, is applied to two PI servos (LB-1005, Newfocus for fast feedback and D2-125, Vescent for slow feedback) after being digitally low-pass and high-pass filtered. The outputs of each servo are then fedback to a voltage-controlled oscillator ($VCO_1$ in Fig. 3(a) that drives the $AOFS_1$) and a lead zirconate titanate (PZT) transducer for fast and slow frequency corrections, respectively.

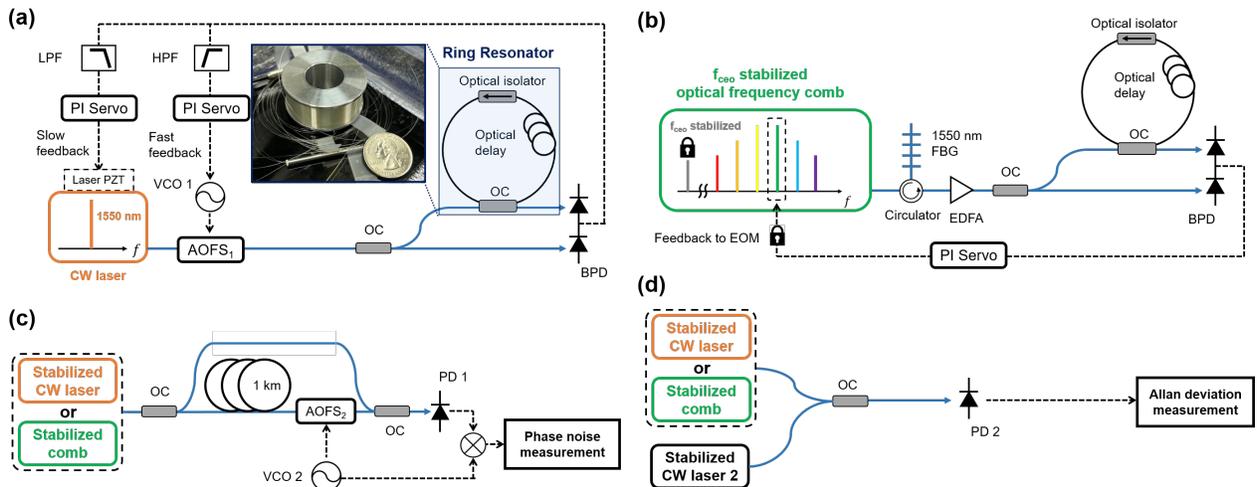

Fig. 3. Schematic of laser stabilization and noise measurement. (a) Schematic of 1550 nm fiber CW laser stabilization based on all-fiber ring-resonator platform. (b) Schematic of 1550 nm comb-line stabilization using the identical setup. (c) Measurement setup of frequency noise of LUT (stabilized CW laser or comb-line) using self-heterodyne method with 1-km fiber delay-line. (d) Measurement setup of Allan deviation by counting beat frequency of LUT and independently stabilized CW laser. AOM, acousto-optic modulator; OC, optical coupler; LPF, low-pass filter; HPF, high-pass filter; VCO, voltage-controlled oscillator; AOFS, acousto-optic frequency shifter; PD, photodiode; BPD, balanced photodiode; EDFA, Erbium-doped fiber amplifier; FBG, fiber bragg grating; EOM, electro-optic modulator.

Fig. 3(b) shows the schematic using identical ring-resonator setup to stabilize the comb-line noise of the self-referenced optical frequency comb. The nonlinear amplifying loop mirror-based erbium fiber frequency comb (FC1500-250-ULN, Menlo Systems GmbH) whose repetition rate is ~250 MHz is used as a comb source. The $f_{ceo}$ of the comb is detected by built-in electronics and stabilized by home-built electronics using 50 MHz signal generator locked to Rb clock. After the 1550-nm comb modes are filtered using the 0.7-nm bandwidth fiber Bragg grating (FBG) then amplified by an Erbium-doped fiber amplifier (EDFA), ~6 mW optical signal is incident to the ring resonator, which is in an air-tight chamber. Note that the repetition rate and $f_{ceo}$ of the optical frequency comb should be controlled carefully to get the enough amount of dip in the error signal, since the filtered optical signal is pulse-like signal in the time domain and the period of the pulse should match well to the fiber delay. By careful control of $f_{rep}$ and $f_{ceo}$, we could obtain 10.8% dip in the error signal. When setting the operation point at the steepest slope point, the sensitivity slope was measured to be 9.4 μV/Hz, which was enough to obtain the fiber thermal noise-limited performance. The detected error signal from balanced photodetector is applied to a single PI servo where the output is fedback to the intra-cavity electro-optic modulator (EOM) in the mode-locked oscillator.

The stabilized CW laser and optical frequency comb are characterized in two ways: (i) the out-of-loop frequency noise measurement using a self-heterodyne method with a 1-km fiber delay-line [Fig. 3(c)] [35] and (ii) the Allan deviation measurement by the beat note detection [Fig. 3(d)]. For the self-heterodyne measurement, we built a Mach-Zehnder interferometer with a 1-km-long fiber delay and an AOFS in the long arm for an unbalanced self-heterodyne. To eliminate the impact of vibrations in the out-of-loop measurement, we employed the vibration-insensitive 1-km fiber spool with vibration sensitivity <$10^{-10}$ [1/g] [the vibration sensitivity of which is shown by curve (iv) of Fig. 2(d)]. The interference signal is then photodetected and mixed with the output of $VCO_2$ (that drives $AOFS_2$) to shift the electric signal to the baseband, and subsequently analyzed by a fast Fourier transform (FFT) spectrum analyzer (SR760, Stanford Research Systems). For the Allan deviation measurement, we built another independently stabilized CW laser and obtained the beat note at ~920 MHz and ~90 MHz, respectively, with the first stabilized CW laser and stabilized comb. The frequency fluctuation of the beat note was sampled at 1 Hz rate with a frequency counter (SR620, Stanford Research Systems), and the frequency instability was evaluated by the overlapping Allan deviation (for >1 s averaging time).

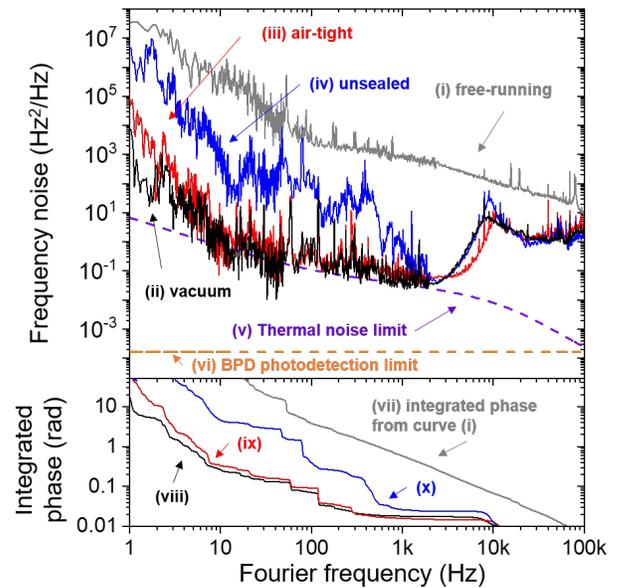

Fig. 4. Frequency noise and integrated phase of stabilized CW laser. (i) frequency noise of the free-running laser; (ii)~(iv) frequency noise of the stabilized laser with reference (ii) in vacuum chamber, (iii) in air-tight chamber and (iv) with no sealing (v) thermal noise limit which is summation of thermo-mechanical and thermo-refractive fiber noise [36], (vi) BPD shot noise limit, (vii)~(x) integrated phase from curve (i)~(iv), respectively.

## V.     Measurement results

### A.    Stabilization of a CW laser

For stabilization of CW laser, when the laser is stabilized by the fiber ring resonator in a low-vacuum chamber (near 1 Torr) the frequency noise level is suppressed by $10^4$ times at Fourier frequency from 100 Hz to 1 kHz [curve (ii) of Fig. 4] compared to the free-running frequency noise [curve (i) of Fig. 4]. In the Fourier frequency range of 10 Hz to 1 kHz, the stabilized frequency noise is limited by the thermal noise of the 100-m fiber delay, which results from thermal fluctuations of the fiber delay length [36, 37]. A large peak near 10 kHz Fourier frequency is the control-loop resonance peak caused by the limited frequency modulation bandwidth of PZT-based actuator. As shown in curve (iii) of Fig. 4, the air-tight case also showed a very similar frequency noise level with the low-vacuum result, only slightly worse in the 1-10 Hz Fourier frequency range. The frequency noise level is increased by $10^2$-$10^4$ times compared to the vacuum case below 1 kHz range [curve (iv) of Fig. 4] when the fiber resonator is placed in an unsealed case, which causes acoustic noise from laboratory fans and air conditioners to be coupled to the frequency noise. The integrated rms optical phase errors of vacuum and air-tight cases are 261 mrad and 335 mrad

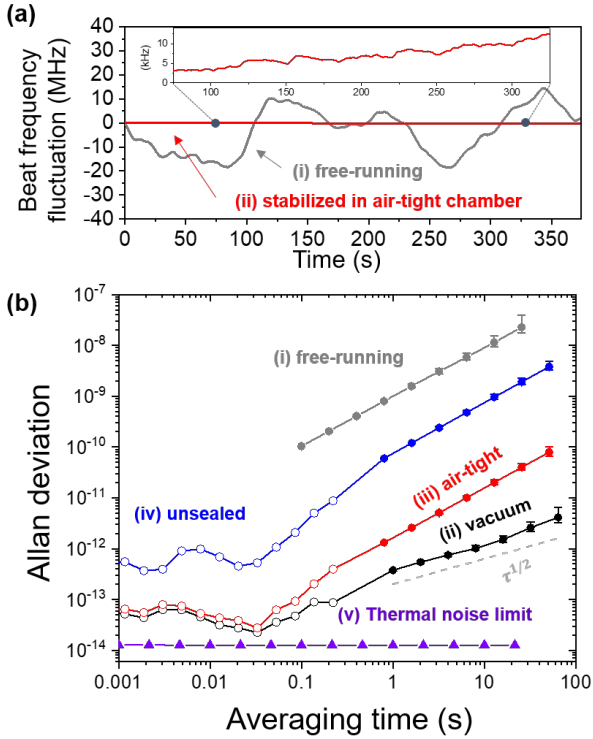

**(a)**

**(b)**

Fig. 5. (a) The frequency fluctuation of beat frequency of (i) two free-running lasers and (ii) two independently stabilized lasers with fiber reference in air-tight chamber. Inset: y-axis-zoomed graph from 75 s to 325 s. (b) Optical frequency instability in terms of overlapping Allan deviation. (i) Free-running laser. (ii) Stabilized laser in vacuum, (iii) in air-tight chamber, and (iv) unsealed chamber (unshaded points - calculated from beat frequency curve (ii)~(iv) in Fig. 4(a); shaded points – calculated from time trace of counted beat frequency). (v) Thermal noise limit converted from curve (v) in Fig. 4.

respectively, when integrated from 10 Hz to 100 kHz. Note that the BPD photodetection limit (shot noise limited) [curve (vi) of Fig. 4] is far below the thermal noise level, demonstrating that the system is not limited by the locking sensitivity of the resonator. The stabilized frequency noise level can reach the fiber thermal noise-limited performance without using the PDH method as well as the vacuum system.

The linewidth can be projected by the $\beta$-separation line method [38] from the measured frequency noise spectrum. The $\beta$-separation linewidth of the CW laser is reduced from 18 kHz (free-running) to 77.4 Hz (stabilized in low vacuum). For the air-tight case, the linewidth is reduced to 309.7 Hz (CW) and 526.8 Hz (comb), which is worse than the vacuum case due to the significantly worse frequency noise spectra below few Hz offset frequency range (see curves (ii) and (iii) in Fig. 4). Note that, for our system, the fundamental limit in linewidth is 6.9 Hz, which is determined by the thermal noise of 100-m fiber.

Fig. 5(a) shows the time trace of the beat frequency of two CW lasers, before and after the independent stabilization in an air-tight chamber. The frequency fluctuation (standard deviation) of the free-running lasers beat note is 7.4 MHz [curve (i) in Fig. 5(a)], while the frequency fluctuation of the beat note of the stabilized lasers is 97 kHz [curve (ii) in Fig.

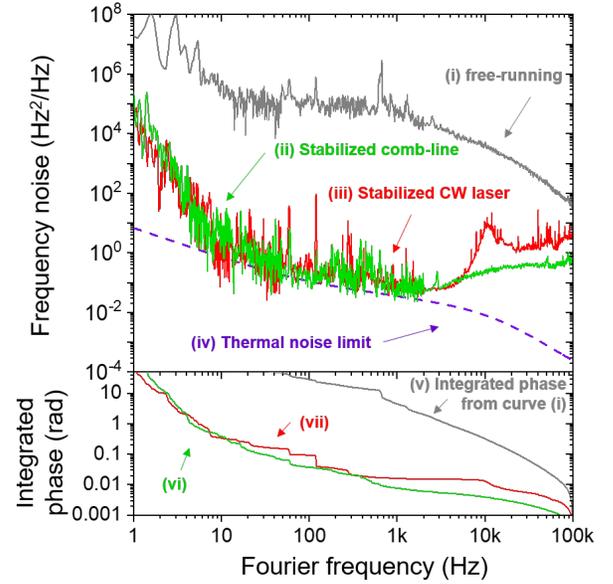

Fig. 6. Frequency noise and integrated phase of stabilized comb-line. (i) frequency noise of the free-running comb-line; (ii) frequency noise of the stabilized comb-line with fiber reference in air-tight chamber. (iii) frequency noise of the stabilized CW laser with fiber reference in air-tight chamber (same data as curve (iii) in Fig. 4) for comparison. (iv) thermal noise limit which is summation of thermo-mechanical and thermo-refractive fiber noise. (v)~(vii) integrated phase from curve (i)~(iii), respectively.

5(a)] over 6 minutes. The time traces (i) and (ii) in Fig. 5(a), and time trace for vacuum and unsealed environment are converted to the frequency instability using overlapping Allan deviation as shown shaded points in curves (i) – (iv) in Fig. 5(b). The frequency instability in air-tight and unsealed chamber diverges fast with $\tau$ slope after 1 s due to the thermal drift. Air-tight chamber can partially block the convection, achieving around 100 times improvement compared to the case with unsealed chamber. The frequency instability in vacuum diverges with $\tau^{1/2}$ slope and 3 times improved at 1-s averaging time compared to that of air-tight case, since vacuum can efficiently block the air convection and the thermal drift is due to the conduction through the contact of the chamber. To evaluate the frequency instability for <1 s averaging time, the Allan deviation is computed from the frequency noise curves (ii) – (iv) of Fig. 4(a) and marked with unshaded points in curve (ii) – (iv) of Fig. 5(b). The frequency instability minimally reaches $2.2 \times 10^{-14}$ and $2.7 \times 10^{-14}$ both at 0.03 s averaging time for vacuum and air-tight environment, respectively.

### B. Stabilization of an optical frequency comb

For stabilization of the self-referenced optical frequency comb in air-tight environment, the frequency noise level is suppressed by maximally $10^6$ times compared to the free-running comb-line noise level [curve (i) of Fig. 6]. The level reaches the fiber thermal-noise limited level in the Fourier frequency range of 10 Hz to 1 kHz [curve (ii) of Fig. 6]. The noise level is identical for the CW laser and the comb-line for

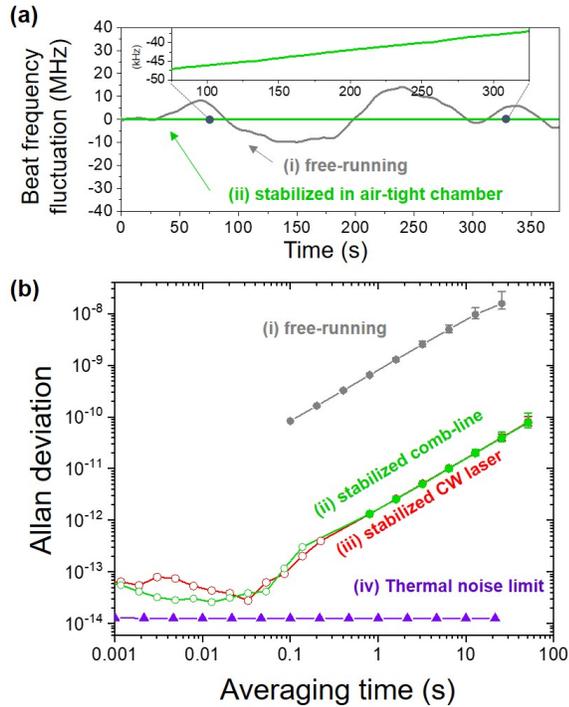

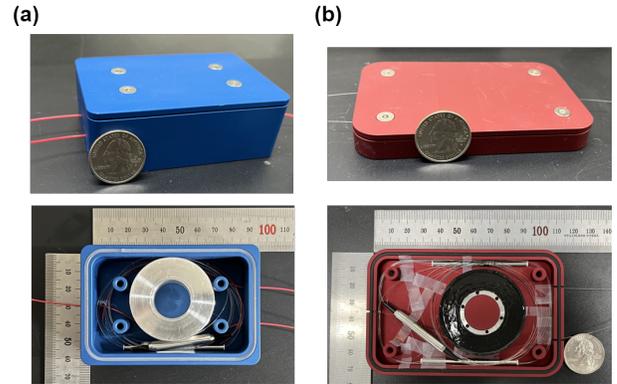

Fig. 8. Packaging of the fiber resonator optics part. (a) A photograph of the packaged fiber resonator using vibration-insensitive spool. (b) A photograph of the packaged fiber resonator using ultracompact (9-mL volume) fiber spool.

Fig. 7. (a) The frequency fluctuation of beat frequency between stabilized CW laser and (i) free-running and (ii) independently stabilized self-referenced optical frequency comb to fiber reference in air-tight chamber. Inset: y-axis-zoomed graph from 75 s to 325 s. (b) Optical frequency instability in terms of overlapping Allan deviation. (i) Free-running comb-line. (ii) Stabilized comb-line in air-tight chamber. (iii) Stabilized CW laser in air-tight chamber (same as curve (iii) in Fig. 5(b)) for comparison and (iv) unsealed chamber (unshaded points -calculated from beat frequency curve (ii)–(iv) in Fig. 4(a); shaded points – calculated from time trace of counted beat frequency). (iv) Thermal noise limit converted from curve (iv) in Fig. 6.

Fourier frequency of less than 1 kHz, meaning that they both follow the length stability of the fiber reference. However, for the comb, the 10 – 20 kHz peak for PZT control is not observed since the comb is stabilized using only the EOM. The locking bandwidth of the feedback is formed near 300 kHz, which is limited by the PI corner of the servo. The integrated rms optical phase error of the system is 380 mrad when integrated from 10 Hz to 100 kHz.

Fig. 7(a) shows the time trace of the beat frequency between 1550-nm comb-line and CW laser, before and after the stabilization of comb-line in air-tight chamber where CW laser stays stabilized in air-tight chamber. The frequency fluctuation of the beat frequency is 5.8 MHz [curve (i) in Fig. 7(a)] and 25 kHz [curve (ii) in Fig. 7(a)] for free-running and stabilized combs, respectively. Same as Fig. 6(a), the shaded points in curve (i) and (ii) is calculated from time trace of beat note in curve (i) and (ii) in Fig. 7(a), and unshaded points in curve (ii) is converted from the frequency noise data curve (ii) in Fig. 6. The curve (iii) is same as curve (iii) in Fig. 5(b) for comparison. The minimum frequency instability for the stabilized comb-line is $2.6 \times 10^{-14}$ at 0.01 s, which is similar to that of the stabilized CW laser. We can conclude that both CW laser and optical frequency comb can follow the stability of the fiber reference.

## VI. Module packaging and vibration test

Finally, we packaged the fiber ring-resonator module in a compact enclosure and examined its mechanical robustness against external vibrations. As shown in Fig. 8(a), the setup is packaged with off-the-shelf optical components (two optical couplers, an optical isolator) and an optical fiber spool in an air-tight aluminum case with dimension of 92 mm × 62 mm × 30 mm. We also packaged the fiber ring-resonator with a miniaturized 100-m-long PM fiber spool made with aluminum with 26-mm diameter (48-mm diameter including fiber), 5-mm height, 9-mL volume and ~$10^{-8}$ vibration sensitivity [curve (iii) of Fig. 2(d)]. The spool and other optical components are packaged in an air-tight aluminum case with dimension of 112 mm × 72 mm × 12 mm [Fig. 8(b)].

To test the impact of external vibrations, we applied vibrations to the packaged module with vibration-insensitive spool [Fig. 8(a)] which stabilized the CW laser, using the shaker (The Modal Shop, 2100E11) with broadband (30 Hz – 1 kHz) vibration spectrum up to $10^{-5}$ [g²/Hz]. The operation of the locking system is monitored by measuring the in-loop error signal as shown in Fig. 9(a). When the vibration is applied, the locking is maintained well thanks to the low vibration sensitivity of the spool, but slightly with worse error signal. Compared to the measured frequency noise without vibrations [curve (ii) of Fig. 9(c)], the frequency noise becomes worse when the vibration level is increasing with the level plotted in Fig. 9(b). The frequency noise levels [curves (iii) of Fig. 9(c)] are measured when the vibration level of the corresponding color in Fig 9(b) is applied. The measured frequency noise with vibration can be easily estimated by multiplying the measured vibration sensitivity [curve (i) of Fig. 2(d)] by the measured vibration spectrum [curve (iii) of Fig. 9(b)]. Note that when the vibration is applied with level of ~$10^{-5}$ [g²/Hz], the frequency noise of the stabilized laser is comparable to that of the free-running laser, which means that the vibration level lower than ~$10^{-5}$ [g²/Hz] should be applied to the system in order to take the advantage from the stabilization system.

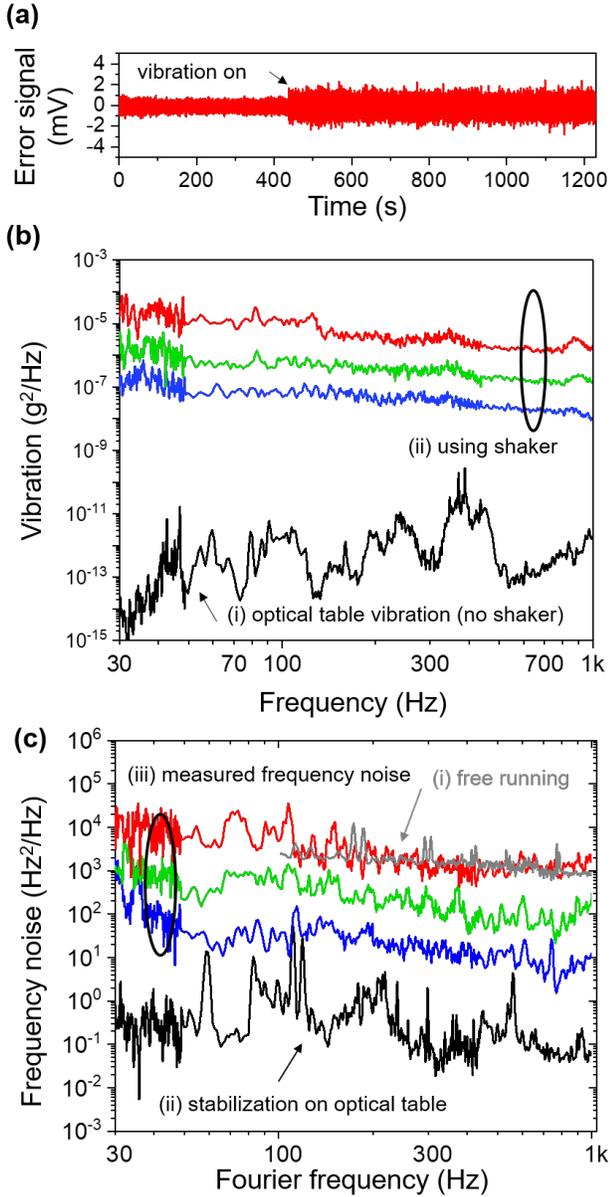

Fig. 9. The vibration test results for packaged setup. (a) The error signal before and after the $10^{-5}$ [g²/Hz]-level vibration is applied. (b) Vibration spectrum of (i) optical table vibration and (ii) shaker with different vibration strength of $10^{-5}$ [g²/Hz], $10^{-6}$ [g²/Hz] and $10^{-7}$ [g²/Hz] for red, green and blue curves, respectively. (c) frequency noise of the stabilized laser using the packaged device: (i) free-running, (ii) stabilized on optical table; (iii) measured frequency noise with vibration plotted in (b).

## VII. Comparison with other methods

In this section, we present a comparison of our work's performance and key characteristics compared to other compact CW laser stabilization platforms as shown in Fig. 10 and Table 1. We mainly compared with compact Fabry-Perot cavities [15, 16, 17], integrated chip-scale resonators [18, 19], bulk dielectric resonators [20], and optical fiber-based SBS [23]. As shown in Fig. 10, our frequency noise PSD and Allan deviation results are among the best; only the recent compact ULE Fabry-Perot cavity with active temperature control within a high vacuum ($10^{-5}$ Pa) chamber environment [16] has a better result. Please note that our method does not necessitate active temperature control and a high vacuum environment, and it can still attain $10^{-14}$-level instability within 0.1 s.

The size, weight, and power consumption (SWaP) is another important metric for assessing different stabilization methods. Because different systems have vastly different operating conditions, making a fair, direct comparison is difficult. For this reason, instead of making a direct SWaP comparison between different systems, we made a comparison based on other metrics using available information. Table 1 shows the input signal coupling, volume/size of the reference part, and the necessity of vacuum chamber and active temperature control for each method, of which are important aspects of SWaP and the practical usability outside laboratory environment. This table shows the cons and pros of each method in terms of size/weight and mechanical robustness. For example, while the compact

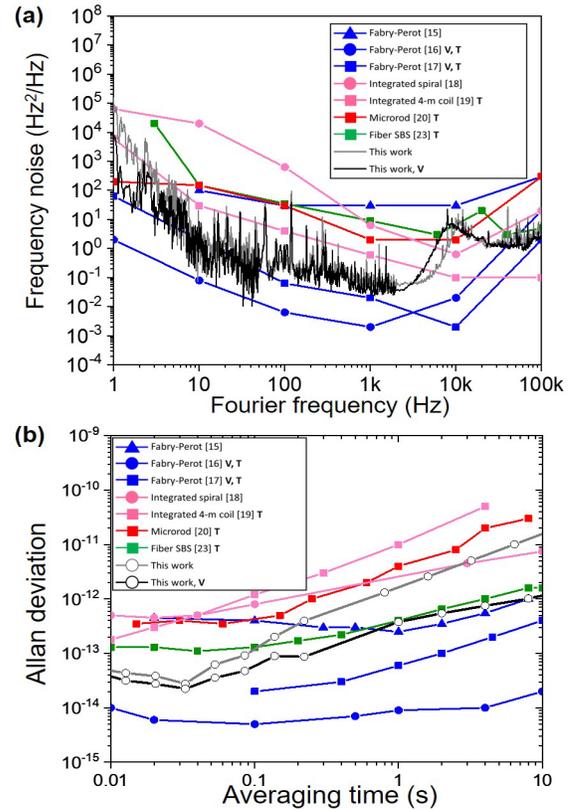

Fig. 10. Performance comparison of various compact state-of-the-art CW laser stabilization methods to this work. (a) Frequency noise PSD. (b) Allan deviation. The V and T symbols in the legend indicate the use of vacuum chamber and active temperature compensation, respectively.

Fabry-Perot cavity's volume itself is small with good stability performance even below $10^{-14}$, it is alignment sensitive (due to free-space coupling) and often require high vacuum and temperature control. The integrated chips can provide the smallest volume and size, but the performance is often limited to the $10^{-13}$ level. Our method can provide the $10^{-14}$-level instability from a mechanically robust all-fiber coupling, relatively small volume, and without vacuum or active temperature control. While we used table-top equipment for the PI servo and the balanced photodetector in this study, they can be replaced by circuit board-based components (e.g., Koheron PI200 and PD10B). Assuming the use of such circuit board-based components in conjunction with our fiber-based reference unit (as shown in Fig. 8(a)), the total stabilization system can be implemented with a fairly low SWaP (<300 mL volume, <300 g weight and ~3 W electric power consumption) for stabilizing CW lasers and frequency combs with $10^{-14}$-level instability.

## VIII. Conclusion

In summary, we have demonstrated a simple, compact and alignment-free laser stabilization module based on an all-fiber ring-resonator and balanced photodetection, which is highly suitable for the field applications outside laboratory environment. The optical frequency noise of a 1550-nm CW laser and comb-line of the self-referenced optical frequency comb is stabilized to the thermal-noise-limit of the used fiber delay, which resulted in the minimum frequency instability of $2.7 \times 10^{-14}$ and $2.6 \times 10^{-14}$ within 0.03-s and 0.01-s averaging time for CW laser and comb, respectively. This stabilization system requires an air-tight atmosphere to accomplish thermal noise-limited performance; a vacuum environment is not required. We could further reduce the vibration-induced frequency noise by designing a vibration-insensitive spool with ~$10^{-10}$ [1/g]-level vibration sensitivity. The fiber optic part could be packaged in a palm-sized aluminum case using either a vibration insensitive spool or an ultracompact 9-mL spool. The performance of stabilization can be further enhanced by increasing the fiber delay length and/or lowering the loss of the fiber ring resonator.

## Acknowledgement

This work is supported by National Research Council of Science and Technology (NST) of Korea (CAP22061-000), Institute for Information and Communications Technology Promotion (IITP) of Korea (RS-2023-00223497), and National Research Foundation (NRF) of Korea (2021R1A2B5B03001407).

## Author declaration

### Conflict of interest

IJ, CA, WJ and JK: Korea Advanced Institute of Science and Technology (P)

### Author Contributions

Igju Jeon: Conceptualization (equal); Data curation (lead); Formal analysis (equal); Investigation (equal); Visualization (lead); Writing – original draft (lead); Writing – review & editing (equal). Changmin Ahn: Conceptualization (equal); Formal analysis (equal); Investigation (equal). Chankyu Kim: Methodology (equal). Seongmin Park: Writing – review & editing (equal). Wonju Jeon: Methodology (equal), Formal analysis (equal). Lingze Duan: Formal analysis (equal). Jungwon Kim: Conceptualization (equal); Formal analysis (equal); Funding acquisition (lead); Project administration (lead), Supervision (lead); Writing – review & editing (equal).

**Table 1. Comparison of various compact state-of-the-art CW laser stabilization platforms to this work**

| Optical reference | Coupling method | Reference volume/size | Vacuum environment | Temperature control | Minimum Allan deviation |
|---|---|---|---|---|---|
| Fabry-Perot [15] | Free-space | 3.2 mL | X | X | $2.5 \times 10^{-13}$ at 1 s |
| Fabry-Perot [16] | Free-space | 8 mL | O | O | $6 \times 10^{-15}$ at 0.1-1 s |
| Fabry-Perot [17] | Free-space | 5.2 mL | O | O | $2 \times 10^{-14}$ at 0.6 s |
| Integrated spiral resonator [18] | Tapered fiber | Chip-scale | X | X | $5.5 \times 10^{-13}$ at 400 µs |
| Integrated coil [19] | Bus-coupling | Chip-scale | X | O | $1.8 \times 10^{-13}$ at 0.01 s |
| Microrod [20] | Tapered fiber | 6-mm diameter | X | O | $3 \times 10^{-13}$ at 0.01-0.1 s |
| Fiber SBS [23] | All-fiber | 50-mm diameter | X | O | $1 \times 10^{-13}$ at 0.01-0.1 s |
| This work | All-fiber | 9 mL | X | X | $2.7 \times 10^{-14}$ at 0.03 s |

**Data availability**

Data underlying the results presented in this paper may be obtained from the authors upon reasonable request.

**Figure captions**

**Fig. 1.** (a) Configuration of the fiber ring resonator with balancing port. (b) Theoretical transmittance of (i) ring resonator port and (ii) balancing port. The transmittance function for the ring resonator port is calculated when the fiber delay is 100 m long and the cavity loss is 0.5 dB.

**Fig. 2.** Vibration-insensitive spool design and measured vibration sensitivity of various spools. (a) Design of a vibration-insensitive optical fiber spool. The fiber (cross section as small circles) is wound in the direction marked by the red dashed-line. (b) Visualization of local vibration sensitivity with vibration at the bottom of the spool. (c) The local vibration sensitivity (i) with groove and (ii) without groove. (d) The vibration sensitivity of various types of 100-m PMF fiber spools and a 1-km SMF fiber spool. (i) designed vibration-insensitive spool; (ii) typical 15-cm diameter plastic spool; (iii) miniaturized (26-mm diameter) aluminum spool without grooves; (iv) vibration-insensitive 1-km SMF spool.

**Fig. 3.** Schematic of laser stabilization and noise measurement. (a) Schematic of 1550 nm fiber CW laser stabilization based on all-fiber ring-resonator platform. (b) Schematic of 1550 nm comb-line stabilization using the identical setup. (c) Measurement setup of frequency noise of LUT (stabilized CW laser or comb-line) using self-heterodyne method with 1-km fiber delay-line. (d) Measurement setup of Allan deviation by counting beat frequency of LUT and independently stabilized CW laser. AOM, acousto-optic modulator; OC, optical coupler; LPF, low-pass filter; HPF, high-pass filter; VCO, voltage-controlled oscillator; AOFS, acousto-optic frequency shifter; PD, photodiode; BPD, balanced photodiode; EDFA, Erbium-doped fiber amplifier; FBG, fiber bragg grating; EOM, electro-optic modulator.

**Fig. 4.** Frequency noise and integrated phase of stabilized CW laser. (i) frequency noise of the free-running laser; (ii)~(iv) frequency noise of the stabilized laser with reference (ii) in vacuum chamber, (iii) in air-tight chamber and (iv) with no sealing (v) thermal noise limit which is summation of thermo-mechanical and thermo-refractive fiber noise [36], (vi) BPD shot noise limit, (vii)~(x) integrated phase from curve (i)~(iv), respectively.

**Fig. 5.** (a) The frequency fluctuation of beat frequency of (i) two free-running lasers and (ii) two independently stabilized lasers with fiber reference in air-tight chamber. Inset: y-axis-zoomed graph from 75 s to 325 s. (b) Optical frequency instability in terms of overlapping Allan deviation. (i) Free-running laser. (ii) Stabilized laser (ii) in vacuum, (iii) in air-tight chamber, and (iv) unsealed chamber (unshaded points - calculated from beat frequency curve (ii)~(iv) in Fig. 4(a); shaded points – calculated from time trace of counted beat frequency). (v) Thermal noise limit converted from curve (v) in Fig. 4.

**Fig. 6.** Frequency noise and integrated phase of stabilized comb-line. (i) frequency noise of the free-running comb-line; (ii) frequency noise of the stabilized comb-line with fiber reference in air-tight chamber. (iii) frequency noise of the stabilized CW laser with fiber reference in air-tight chamber (same data as curve (iii) in Fig. 4) for comparison. (iv) thermal noise limit which is summation of thermo-mechanical and thermo-refractive fiber noise. (v)~(vii) integrated phase from curve (i)~(iii), respectively.

**Fig. 7.** (a) The frequency fluctuation of beat frequency between stabilized CW laser and (i) free-running and (ii) independently stabilized self-referenced optical frequency comb to fiber reference in air-tight chamber. Inset: y-axis-zoomed graph from 75 s to 325 s. (b) Optical frequency instability in terms of overlapping Allan deviation. (i) Free-running comb-line. (ii) Stabilized comb-line in air-tight chamber. (iii) Stabilized CW laser in air-tight chamber (same as curve (iii) in Fig. 5(b)) for comparison and (iv) unsealed chamber (unshaded points -calculated from beat frequency curve (ii)~(iv) in Fig. 4(a); shaded points – calculated from time trace of counted beat frequency). (iv) Thermal noise limit converted from curve (v) in Fig. 6.

**Fig. 8.** Packaging of the fiber resonator optics part. (a) A photograph of the packaged fiber resonator using vibration-insensitive spool. (b) A photograph of the packaged fiber resonator using ultracompact (9-mL volume) fiber spool.

**Fig. 9.** The vibration test results for packaged setup. (a) The error signal before and after the $10^{-5}$ [$g^2$/Hz]-level vibration is applied. (b) Vibration spectrum of (i) optical table vibration and (ii) shaker with different vibration strength of $10^{-5}$ [$g^2$/Hz], $10^{-6}$ [$g^2$/Hz] and $10^{-7}$ [$g^2$/Hz] for red, green and blue curves, respectively. (c) frequency noise of the stabilized laser using the packaged device: (i) free-running, (ii) stabilized on optical table; (iii) measured frequency noise with vibration plotted in (b).

**Fig. 10.** Performance comparison of various compact state-of-the-art CW laser stabilization methods to this work. (a) Frequency noise PSD. (b) Allan deviation. The V and T symbols in the legend indicate the use of vacuum chamber and active temperature compensation, respectively.